\documentclass[showpacs,preprintnumbers,superscriptaddress,amsmath,reprint,aps,prd]{revtex4-1}
\usepackage{comment}
\usepackage{slashed}
\usepackage{latexsym}
\usepackage{graphics}
\usepackage{bm}
\usepackage{graphicx} 
\usepackage{color}
\usepackage{amsmath}
\usepackage{amssymb}
\usepackage{feynmp-auto}
\usepackage{slashed}
\usepackage{color,hyperref}
\hypersetup{colorlinks,breaklinks,
            linkcolor=blue,urlcolor=blue,
            anchorcolor=blue,citecolor=blue}

\usepackage{tikz}
\usetikzlibrary{shapes.geometric,decorations.fractals,shadows,shapes.multipart,arrows,snakes,positioning,arrows}
\usetikzlibrary{decorations.pathmorphing,decorations.markings}
\usetikzlibrary{calc,intersections}

\newcommand{\be}{\begin{equation}}
\newcommand{\ee}{\end{equation}}
\newcommand{\bea}{\begin{eqnarray}}
\newcommand{\eea}{\end{eqnarray}}

\newcommand{\Co}{$^{60}$Co}
\newcommand{\Ni}{$^{60}$Ni}
\newcommand{\Na}{$^{24}$Na}

\begin{document}
\title{The GAmmas from Nuclear Decays Hiding from Investigators (GANDHI) Experiment}
\author{Giovanni Benato}
\affiliation{Department of Physics, University of California, Berkeley, CA 94720, USA}
\affiliation{Nuclear Science Division, Lawrence Berkeley National Laboratory, Berkeley, CA 94720, USA}
\author{Alexey Drobizhev}
\affiliation{Nuclear Science Division, Lawrence Berkeley National Laboratory, Berkeley, CA 94720, USA}
\author{Surjeet Rajendran}
\affiliation{Berkeley Center for Theoretical Physics, Department of Physics,
University of California, Berkeley, CA 94720, USA}
\author{Harikrishnan Ramani}\email{hramani@berkeley.edu}
\affiliation{Berkeley Center for Theoretical Physics, Department of Physics,
University of California, Berkeley, CA 94720, USA}
\affiliation{Theoretical Physics Group, Lawrence Berkeley National Laboratory, Berkeley, CA 94720}
\begin{abstract}
We propose a high statistics experiment to search for invisible decay modes in nuclear gamma cascades. A radioactive source (such as  \Co\ or \Na) that triggers gamma cascades  is placed in the middle of a large, hermetically sealed scintillation detector, enabling photon identification with high accuracy. Invisible modes are identified by establishing the absence of a photon in a well-identified gamma cascade. We propose the use of fast scintillators with nanosecond timing resolution, permitting event rates as high as $10^{7}$ Hz. Our analysis of the feasibility of this setup indicates that branching fractions as small as $10^{-12} - 10^{-14}$ can be probed. 
This experimental protocol benefits from the fact that a search for invisible modes is penalized for weak coupling only in the production of the new particle. If successfully implemented, this experiment is an exquisite probe of particles with mass below $\sim$4 MeV that lie in the  poorly constrained supernova ``trapping window'' that exists between 100 keV - 30 MeV. Such particles have been invoked as mediators between dark matter and nucleons, explain the proton radius and $(g-2)_{\mu}$ anomalies and potentially power the shock wave in type II supernovae. The hadronic axion could also be probed with modifications to the proposed setup.
\end{abstract}
\maketitle

\section{Introduction}
Weakly-coupled particles with mass below the MeV scale arise in a number of extensions of the Standard Model. They are natural ingredients of the dark sector---either as dark matter candidates or as mediators to the dark matter \cite{Green:2017ybv}. They may also play a role in solving the gauge hierarchy and vacuum energy problems through cosmological evolution and could conceivably be responsible for powering the shock wave necessary to trigger type II supernovae~\cite{Hidaka:2007se,surjeetforth}.  High energy colliders do not have the statistical sensitivity to search for these particles---while they operate well above the energy threshold necessary to produce these particles, colliders do not have the luminosity to overcome the small coupling of these particles. In this paper, we propose a new method for a high statistics search for such particles. 

We aim to search for invisible particles in nuclear gamma decays
using the following protocol.  We place a radioactive source in the
middle of a hermetically sealed detector. The source is chosen so
that its decay triggers a gamma cascade in the daughter nucleus. The
photons in the cascade are accurately counted and their energies 
measured. An invisible decay mode would manifest itself as a missing 
photon in this count.  As a concrete example, consider  \Co\ which 
undergoes a $\beta$ decay to \Ni\ (see Figure~\ref{schemeCo}). This 
decay (with high probability) places \Ni\ at a 2.51-MeV excited state, 
which de-excites into another excited state by emitting a 1.17-MeV gamma. 
This excited state subsequently decays to the ground state by emitting a 
1.33-MeV gamma. A decay in which a 1.17-MeV gamma is not followed by a 
1.33-MeV gamma would indicate the presence of a new light, weakly-coupled
particle. The ready availability of the \Co\ source allows a high
statistics search, potentially enabling a probe of weakly-coupled
particles. A major advantage of this scheme is that the search for an
invisible decay only suffers the cost of the weak coupling in the
production of the particle, unlike other proposed schemes that rely
on production and subsequent decay/scattering of the light particle.
While there are experiments such as NA64~\cite{Banerjee:2016tad}, LDMX~\cite{Mans:2017vej}, BDX~\cite{Battaglieri:2014qoa} that directly
probe missing momentum/energy produced due to new particles coupled
to electrons, there have not been similar efforts to probe baryonic
interactions. 
Hunting for new forces in nuclear decays has been considered in early Higgs searches and axion searches~\cite{Calaprice:1979pe, Zehnder:1981qn,Cavaignac:1982ek, Alekseev:1982iq, Lehmann:1982bp, Zehnder:1982bg, Ananev:1983ki,Hallin:1986gh, Avignone:1988bv,Datar:1988ju, DeBoer:1988ts, Dohner:1988uv,Savage:1988rg, Bini:1989zq, Asanuma:1990rm,deBoer:1990yg,Hicks:1992ie,Tsunoda:1995at,deBoer:1997mr}, where visible decays were nonetheless required. There has been a recent revival of similar ideas, albeit with detecting the final state involving an electron coupling in~\cite{Izaguirre:2014cza,Pospelov:2017kep,Kozaczuk:2017per}. There have been few attempts at looking for the QCD axion in invisible nuclear decays in~\cite{Minowa:1993mw,Derbin:1997kt,Derbin:2002qq} albeit at lower energies and statistics.

In this paper, we analyze the experimental strategies necessary to implement the above scheme in a high statistics environment. First, it is essential that the experiment observes all the photons produced in the decay. We aim to accomplish this using a sufficiently large scintillation detector so that there are enough radiation lengths in the detector to contain all the decays. Second, the experiment needs to avoid pile up of events while maintaining high statistics. We aim to accomplish this goal by using plastic scintillators that have short ($\sim$ 10 ns) decay times and employing modular construction to separate events spatially. Finally, the scheme would have to distinguish the signal from a variety of systematic backgrounds such as mis-identification of the gammas, potential confusion introduced by soft Compton scattering and population of other nuclear levels by the decaying source.

The feasibility of such an experiment is the focus of this paper. Since the missing particles are produced in the decays of nuclear isomers, the experiment is maximally sensitive to particles coupled to baryons.  Our results indicate that this scheme has the potential to probe invisible branching fractions $\sim 10^{-12} - 10^{-14}$. This is of significant interest:  there are poor limits on particles with mass $\gtrapprox 100$ keV that couple to baryons \cite{Green:2017ybv}. Several dark matter experiments are presently under development to search for interactions between the dark matter and the Standard Model mediated by particles around this mass scale. Particles in this parameter space have also been invoked to explain the proton radius and $\left(g-2\right)_{\mu}$ anomalies. Moreover, such light particles can be produced in type II supernova and their cosmic populations can potentially be discovered in current dark matter detectors. A light, weakly-coupled particle in this scale could significantly affect the dynamics of type II supernovae, potentially resolving long standing puzzles associated with the production of shock waves necessary to trigger such explosions. An experiment that can probe invisible branching fractions $\sim 10^{-12} - 10^{-14}$ thus has significant phenomenological implications.

We start by presenting a simple toy model in Section~\ref{sec5}, where we review current bounds and identify the experimentally accessible parameter space. 
Following this, we describe the experiment in greater detail in  Section~\ref{sec:expconcept}. Section~\ref{sec:selection} deals with the event selection. Sensitivity and backgrounds are treated in Section~\ref{sec:sensitivity},
while the technological challenges and further improvements are presented
in Section~\ref{sec:challenges}. Finally, we conclude in Section~\ref{sec6}.

\section{A Toy Model \label{sec5}}



Consider the operator: 
\begin{equation}
\mathcal{L} = g_p \phi \bar{p}p
\end{equation}
describing the interactions of a light scalar $\phi$ with protons $p$. Such a scalar is a popular way to couple nucleons and dark matter.  $\phi$ can be emitted during nuclear decays and can be probed by this proposed experiment. Taking this as our benchmark model, we quote the sensitivity of our experiment in terms of the coupling $g_p$. 

There are a variety of constraints on $\phi$.  Major constraints arise from astrophysics; cooling of stars / supernovae from energy carried away by this scalar. These are treated in detail in various texts \cite{Raffelt:1996wa} and we summarize the major results here. There are strong constraints on $\hat{g}_p$ from horizontal branch stars and red giants  when the mass of $\phi$ is less than 100 keV. Above 100 keV, there are limits from the cooling of SN1987A up to 100 MeV. However, for moderately large couplings, $\phi$ is trapped in the supernova and does not contribute to cooling. This trapping window is a prime target for the proposed experiment. 

In addition to these astrophysical constraints, there are direct constraints from terrestrial experiments on coupling to nucleons. These are summarized in  \cite{Liu:2016qwd} and are relatively weak. Additional constraints can be placed on this scenario from UV-completing this model. These are somewhat model dependent. For example, this nucleon coupling can be generated via  heavy quark couplings or through gluons (via the operator $\phi GG$). Limits from Kaon decays can set limits on these models. These were considered in detail in \cite{Knapen:2017xzo}. Here we instead consider coupling to light quarks, which are not as constrained.  

Starting with $y_q\phi \bar{q}{q}$ leads both to a meson coupling and nucleon coupling,
\begin{equation}
\mathcal{L} \supset \frac{y_q}{m_q} \phi(m_\pi^2\pi^+ \pi^- + f^q_N m_N\bar{N}N)
\end{equation}
where the $f^q_N$ for light quarks are tabulated in \cite{Hoferichter:2017olk}.
Here $g_N = \frac{g_q}{m_q} f^q_N m_N$.
The former term in the effective Lagrangian results in a new decay channel $K^+ \rightarrow \pi^+ \phi$. The branching fraction in the $m_K \gg m_{\pi} \gg m_{\phi}$ limit is given by:
\begin{equation}
\frac{\Gamma_{K\rightarrow \pi \phi}}{\Gamma_{K\rightarrow \mu \nu_{\mu}}}=\frac{3y_q^2 f_{\pi}^2  m_{\pi}^4}{4m_K^2 m_\mu^2
   m_q^2}
\end{equation}

The branching ratio for invisible decays of charged Kaons is constrained to $Br(K \rightarrow \pi \phi) \le 1.7. 10^{-10}$ {}. This then sets a limit $g_N \le 4. 10^{-5}$. 



\subsection{Reach}
Our analysis of the experimental reach suggests a sensitivity to invisible decay modes with a branching ratio $\sim 10^{-12} - 10^{-14}$. In this sub-section, we describe the conversion between this experimental sensitivity and the coupling $g_p$. 

This particular computation strictly applies to couplings to protons and ignores the coupling to neutrons because of the ease of porting known photon matrix elements to those of $\phi$. The estimate of the branching fraction also requires knowledge of the specific nuclear transitions. In our proposed experiment, we consider two promising radioactive sources (see Section~\ref{sec:expconcept} for details): \Co \, and \Na. In both these cases the relevant gamma transitions are $E_2$ transitions. 

\begin{figure*}[htbp]
\centering
\includegraphics[width=2\columnwidth,keepaspectratio]{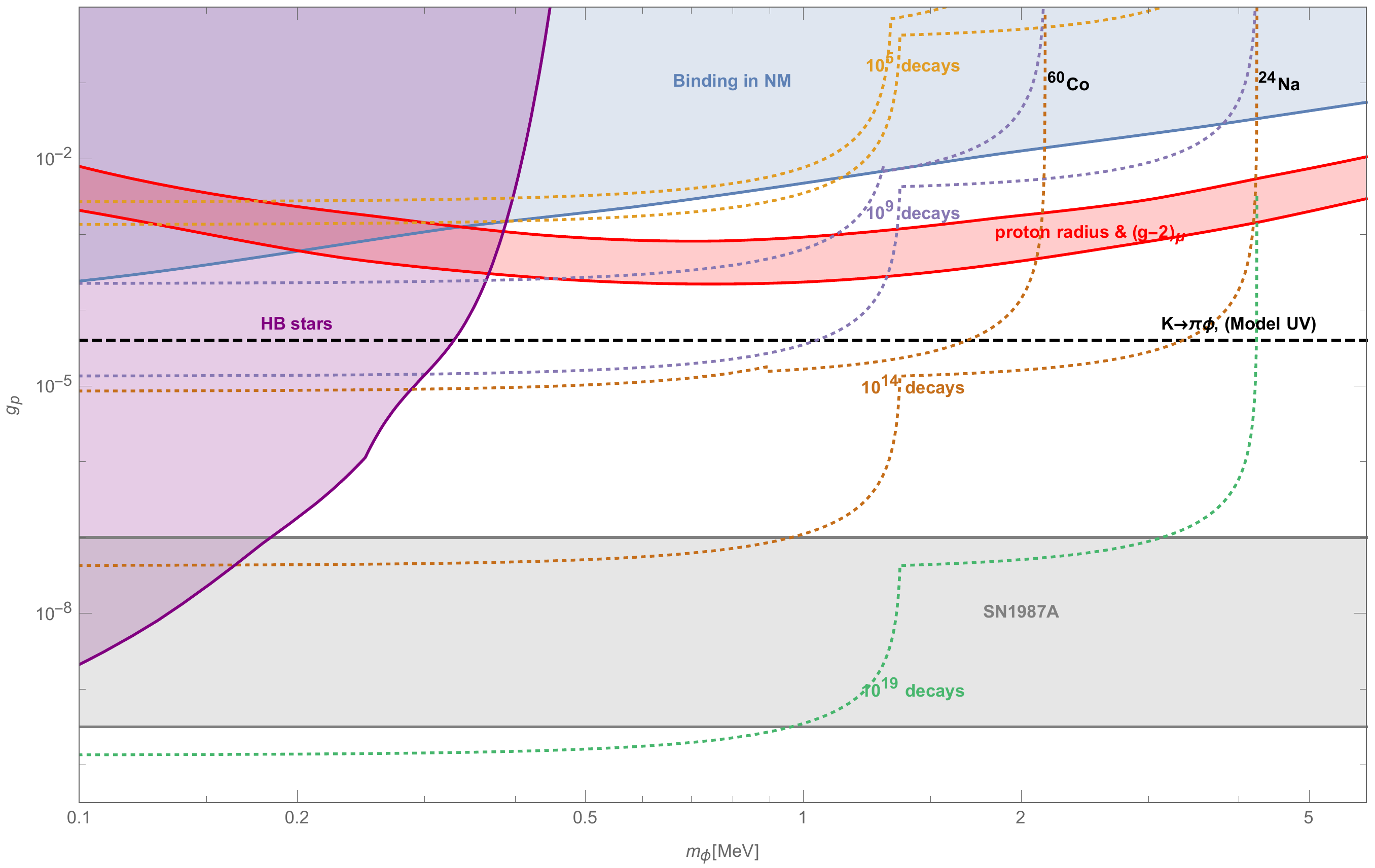}
\caption{Reach for $ ^{60} Co$ and $ ^{24} Na$ experimental proposals for $10^{5}$, $10^{10}$, $10^{14}$ and $10^{19}$ decaying mother nuclei. Also shown are direct limits from binding in Nuclear matter \cite{Liu:2016qwd}, limits from cooling in SN1987A  and Horizontal Branch stars \cite{Raffelt:1996wa} and indirect limits from  meson decays in a UV complete model (refer text). Also shown in red is the region that could explain the proton radius and muon g-2 puzzles simultaneously\cite{Liu:2016qwd}.}
\label{expreach}%
\end{figure*}

For an $E_2$ transition, the quadrupole Hamiltonian that is induced by this Yukawa coupling is(see for e.g. \cite{Pospelov:2017kep}): 
\begin{equation}
H^{\phi}_{\text{int}}=g_p R_p^i R_p^j \nabla_i \nabla_j \phi(k)\mathrm{~,}
\end{equation}
\noindent where $\phi(k)$ is the free-particle wave-function. Comparing this to 
\begin{equation}
H^{\gamma}_{\text{int}}=e R_p^i R_p^j \nabla_i  \epsilon_j \mathrm{~,}
\end{equation}
\noindent notice that for a massive scalar, the momentum $k=\sqrt{\omega^2-m_{\phi}^2}$, where $\omega$ is the energy gap of the transition. Putting this all together,
\begin{equation}\label{rateratio}
\frac{\Gamma(\phi)}{\Gamma_{\gamma,E_2}} \sim \frac{1}{2} \left(\frac{g_p}{e}\right)^2 \left(1-\frac{m_{\phi}^2}{\omega^2}\right)^{\frac{5}{2}} \mathrm{~.}
\end{equation}
Assuming a 100 \% efficiency in photon detection, we plot the reach for $10^{5}$,$10^{10}$ and $10^{14}$ decaying mother nuclei. These roughly correspond to $10^{-2}$ Hz, $10^3$ Hz, $10^7$ Hz triggering frequency for a 1 year run. The reach is plotted in Figure~\ref{expreach}
for experiments using \Co\ and \Na. If this sensitivity is successfully attained, this experiment will probe the entire trapping window of mediators in the mass range 100 keV  to $\sim$ 1.3 MeV. 


\section{Experimental Concept \label{sec:expconcept}}

From Equation~\ref{rateratio} it follows that an experiment searching for the disappearance of a gamma ray
with energy $E_{\gamma}$ is sensitive to a scalar $\phi$ with a mass up to $E_{\gamma}$.
A positive signal would allow the measurement of the coupling term $g_p$, but not of $m_{\phi}$.
The experimental sensitivity goes as the ratio between the decay rate into the dark sector and the standard one:
\begin{equation}
\hat{g_p} = \frac{g_p}{e} \propto \sqrt{ \frac{\Gamma(\phi)}{\Gamma_{\gamma}} } .
\end{equation}
Figure~\ref{expreach} shows two unexplored regions, corresponding to $[10^5,10^{14}]$ decays,
and to $>10^{19}$ decays. While the latter is hardly feasible with current detector technology,
it is possible to probe the first region with one year of livetime,
provided that the experimental apparatus can sustain a count rate of $\sim10$~MHz
and has a containment efficiency $\varepsilon_c\geq\left(1-10^{-14}\right)$.
Such a high efficiency is only reachable with a liquid or solid detector
with a large enough continuous active volume. 

Additional requirements are imposed by the choice of the gamma emitting isotope.
To maximize the accessible parameter space region, we want $E_{\gamma}$ to be large,
above the current HB stars exclusion limits at $\sim200$~keV.
Moreover, the considered decay must offer a clean signature with no intrinsic backgrounds.
Restricting our discussion to $\alpha$ and $\beta$ decaying isotopes only,
we need to be able to detect distinct energy depositions for the $\alpha$/$\beta$
and the daughter gamma ray(s). In practice, any radioactive source has a finite size
and $\alpha$ particles of few MeV have a range of order of tens of nm in average-Z materials.
This strongly suppresses the $\alpha$ detection efficiency for all those atoms
which are not on the surface of the source,
and leads us to the choice of $\beta$ decaying nuclei.
One possibility is to select an isotope that undergoes $\beta$ decay followed by
a single gamma de-excitation of the daughter nucleus, such as $^{137}$Cs.
The signal signature would be an energy deposition compatible
with that of the $\beta$ in a location next to that of the source.
A much more identifiable signature would be that of a $\beta$ decay
followed by two gammas in cascade, as for example in the case of $^{60}$Co.
A signal-like event would be characterized by the $\beta$ energy deposition
in vicinity of the source, and a gamma energy deposition elsewhere in the detector volume.
The distinction between the two energy depositions requires a specific space resolution,
which depends on the energies of the involved particles.
Typically, gamma from nuclear de-excitations are emitted within ps or ns from the original decay
and would be considered in coincidence with it for most detector technologies.
The requirement of a double coincidence within a $O($ns$)$ time window strongly suppresses
random coincidences and background events induced by isotopes decaying in cascade.
On the other hand, the potential presence of intrinsic backgrounds
induced by the source itself and mimicking
the gamma disappearance must be considered in the isotope choice.

The practicality of the source production and usage imposes additional requirements
to the selection of the isotope and of the detector technology.
First of all, the isotope half-life has to be long enough to allow the source
transportation to the experiment site, its insertion in the detector,
and a measurement time sufficient to collect the required statistics.
Thus, isotopes with an half-life $\gtrsim1$~yr are preferable.
Alternatively, we could envisage the repeated production and insertion of the source
in the experimental apparatus, provided that this is close enough to the production site.
Such a choice allows the use of isotopes with half-lifes down to several hours,
but imposes the capability to insert and extract the source in the detector
without affecting its performance, and the availability of a long term dedicated
source production facility, e.g. a beam line.
Furthermore, the necessity of measuring the $\beta$ in a given location
restricts our choice to solid state sources on thin enough materials to minimize the self-absorption.
Finally, isotopes for which a production technology exists with industrial standards are preferable.

\begin{figure}[htbp]
	\begin{tikzpicture}[xscale=1,yscale=1]
    
    \def\d{0.9}
    \def\xr{\d+4.}
    \def\Q{3.8228}
    \def\yfEx{1.335}
    \def\ysEx{2.1586}
    \def\ytEx{2.5058}
    \def\dt{0.5mm}
    
    \node[anchor=east,font=\large] at (0,\Q)(Co){$^{60}$Co};
    \node[anchor=west,font=\footnotesize] at (\d,\Q)(rCo){$0$~MeV, $5+$};
    \node[font=\footnotesize] at (0.5*\d,\Q+0.2) (T12) {$T_{1/2}=5.3$~yr};
	\draw[line width=\dt] (Co) -- (rCo);
    
    \node[anchor=east,font=\large] at (\xr,0)(Ni){$^{60}$Ni};
    \node[anchor=west,font=\footnotesize] at (\xr+\d,0)(rNi){$0$~MeV, $0+$};
    \draw[line width=\dt] (Ni) -- (rNi);
    
    \node[anchor=east] at (\xr,\yfEx)(fEx){};
    \node[anchor=west,font=\footnotesize] at (\xr+\d,\yfEx)(rfEx){$1.33$~MeV, $2+$};
    \draw[line width=\dt] (fEx) -- (rfEx);
    
    \node[anchor=east] at (\xr,\ysEx)(sEx){};
    \node[anchor=west,font=\footnotesize] at (\xr+\d,\ysEx)(rsEx){$2.16$~MeV, $2+$};
    \draw (sEx) -- (rsEx);
    
    \node[anchor=east] at (\xr,\ytEx)(tEx){};
    \node[anchor=west,font=\footnotesize] at (\xr+\d,\ytEx)(rtEx){$2.51$~MeV, $4+$};
    \draw[line width=\dt] (tEx) -- (rtEx);
    
    \draw[-latex] (0.5*\d,\Q) -- (\xr,\ytEx);
    \draw[-latex] (0.5*\d,\Q) -- (\xr,\ysEx);
    \draw[-latex] (0.5*\d,\Q) -- (\xr,\yfEx);
    
    \node[rotate=0,font=\footnotesize] at (2.8,3.35)(b1){$\beta_1$};
    \node[rotate=0,font=\footnotesize] at (2.8,2.8)(b2){$\beta_2$};
    \node[rotate=0,font=\footnotesize] at (2.8,2.2)(b1){$\beta_3$};
    \draw[-latex,decorate,decoration={snake,amplitude=.2mm,segment length=1mm,post length=1mm}] (\xr+0.2,\ytEx) -- (\xr+0.2,\yfEx);
    \draw[-latex,decorate,decoration={snake,amplitude=.2mm,segment length=1mm,post length=1mm}] (\xr+0.6,\ytEx) -- (\xr+0.6,\ysEx);
    \draw[-latex,decorate,decoration={snake,amplitude=.2mm,segment length=1mm,post length=1mm}] (\xr+0.9,\ysEx) -- (\xr+0.9,0);
    \draw[-latex,decorate,decoration={snake,amplitude=.2mm,segment length=1mm,post length=1mm}] (\xr+0.4,\yfEx) -- (\xr+0.4,0);
    \node[anchor=west,font=\footnotesize] at (\xr-6,\yfEx-0.6) (g1){$\beta_1$,$99.88\%$,$0.32$~MeV};
    \node[anchor=west,font=\footnotesize] at (\xr-6,\yfEx-1.0) (g1) {$\beta_2$,$0.001 \%$,$0.67$~MeV};
    \node[anchor=west,font=\footnotesize] at (\xr-6,\yfEx-1.4) (g1) {$\beta_3$,$0.12\%$,$1.49$~MeV};
    \node[anchor=west,font=\footnotesize] at (\xr-6,\yfEx+1) (g1) {$E\gamma_1= 1.17$ MeV};
    \node[anchor=west,font=\footnotesize] at (\xr-6,\yfEx+0.6) (g12) {$E_{\gamma_2} = 1.33$ MeV};
    \node[anchor=west,font=\footnotesize] at (\xr-6,\yfEx+0.2) (g12) {$E_{\gamma_3} = 0.35$ MeV};
    \node[anchor=west,font=\footnotesize] at (\xr-6,\yfEx-0.2) (g12) {$E_{\gamma_4} = 2.16$ MeV};
    \node[anchor=west,font=\footnotesize] at (\xr-0.25,\yfEx+0.4) (g1) {${\gamma_1}$};
    \node[anchor=west,font=\footnotesize] at (\xr-0.1,0.6) (g12) {$\gamma_2$};
    \node[anchor=west,font=\footnotesize] at (\xr+0.5,\yfEx+1) (g1) {${\gamma_3}$};
    \node[anchor=west,font=\footnotesize] at (\xr+0.47,0.6) (g12) {$\gamma_4$};
	\end{tikzpicture}
    \caption{Decay scheme of $^{60}$Co}\label{schemeCo}
\end{figure}
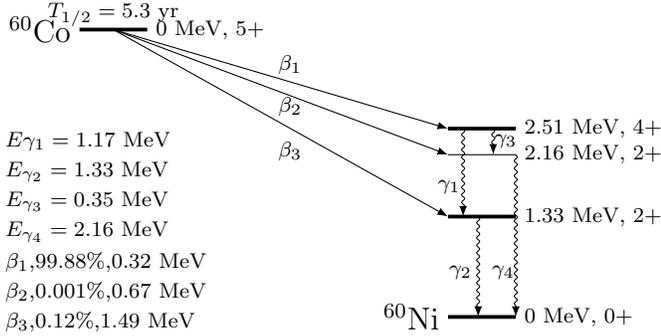

Two isotopes that fulfill most of these criteria are \Co\ and \Na.
With \Co\ (Figure~\ref{schemeCo}), we can search for the disappearance of the $1.33$~MeV gamma.
The signal signature is therefore a twofold energy deposition by the $\beta$ and the $1.17$~MeV gamma.
On the one hand, \Co\ is a commercially available isotope with a half-life that perfectly fits
the live time of a hypothetical experiment. On the other hand, the relatively low end-point
of the $\beta$ spectrum ($0.32$~MeV) and the small difference between the energy of the two gammas
impose strict requirements in terms of energy threshold and resolution.
Furthermore, \Co\ is affected by an intrinsic background in that it has a $0.12\%$ branching ratio
into the $1.33$~MeV excited state of $^{60}$Ni. A signal like event can be detected
if most of the energy is carried away by the anti-neutrino, and if the $1.33$~MeV gamma undergoes a soft-Compton scattering in proximity of the source and is then fully absorbed elsewhere.
The actual importance of this background strongly depends on the detector material
and on the spatial resolution.

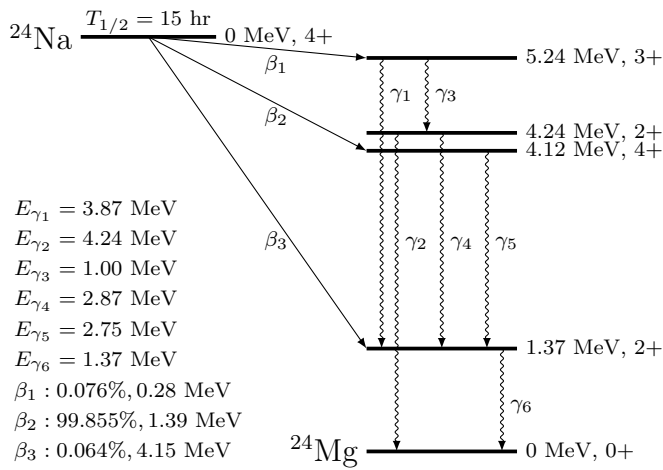
\begin{figure}[htbp]
	\begin{tikzpicture}[xscale=1,yscale=1]
    
    \def\d{1.8}
    \def\dr{2.}
    \def\xr{\d+2}
    \def\Q{5.5157}
    \def\yfEx{1.3687}
    \def\ysEx{4.0}
    \def\ytEx{4.2384}
    \def\yqEx{5.2351}
    \def\dt{0.5mm}
    
    \node[anchor=east,font=\large] at (0,\Q)(Na){$^{24}$Na};
    \node[anchor=west,font=\footnotesize] at (\d,\Q)(rNa){$0$~MeV, $4+$};
    \node[font=\footnotesize] at (0.5*\d,\Q+0.2) (T12) {$T_{1/2}=15$~hr};
	\draw[line width=\dt] (Na) -- (rNa);
    
    \node[anchor=east,font=\large] at (\xr,0)(Mg){$^{24}$Mg};
    \node[anchor=west,font=\footnotesize] at (\xr+\dr,0)(rMg){$0$~MeV, $0+$};
    \draw[line width=\dt] (Mg) -- (rMg);
    
    \node[anchor=east] at (\xr,\yfEx)(fEx){};
    \node[anchor=west,font=\footnotesize] at (\xr+\dr,\yfEx)(rfEx){$1.37$~MeV, $2+$};
    \draw[line width=\dt] (fEx) -- (rfEx);
    
    \node[anchor=east] at (\xr,\ysEx)(sEx){};
    \node[anchor=west,font=\footnotesize] at (\xr+\dr,\ysEx)(rsEx){$4.12$~MeV, $4+$};
    \draw[line width=\dt] (sEx) -- (rsEx);
    
    \node[anchor=east] at (\xr,\ytEx)(tEx){};
    \node[anchor=west,font=\footnotesize] at (\xr+\dr,\ytEx)(rtEx){$4.24$~MeV, $2+$};
    \draw[line width=\dt] (tEx) -- (rtEx);
    
    \node[anchor=east] at (\xr,\yqEx)(qEx){};
    \node[anchor=west,font=\footnotesize] at (\xr+\dr,\yqEx)(rqEx){$5.24$~MeV, $3+$};
    \draw[line width=\dt] (qEx) -- (rqEx);
    
    \draw[-latex] (0.5*\d,\Q) -- (\xr,\yqEx);
    \draw[-latex] (0.5*\d,\Q) -- (\xr,\ysEx);
    \draw[-latex] (0.5*\d,\Q) -- (\xr,\yfEx);
    
    \node[font=\footnotesize] at (2.6,5.15)(b1){$\beta_1$};
    \node[font=\footnotesize] at (2.6,4.45)(b1){$\beta_2$};
    \node[font=\footnotesize] at (2.6,2.80)(b2){$\beta_3$};
    
    \node[anchor=west,font=\footnotesize] at (-1,0.8) (c1){$\beta_1: 0.076\%, 0.28$~MeV};
    \node[anchor=west,font=\footnotesize] at (-1,0.4) (c2){$\beta_2: 99.855\%, 1.39$~MeV};
    \node[anchor=west,font=\footnotesize] at (-1,0.0) (c3){$\beta_3: 0.064\%, 4.15$~MeV};
    
    \draw[-latex,decorate,decoration={snake,amplitude=.2mm,segment length=1mm,post length=1mm}] (\xr+0.2,\yqEx) -- (\xr+0.2,\yfEx);
    \draw[-latex,decorate,decoration={snake,amplitude=.2mm,segment length=1mm,post length=1mm}] (\xr+0.4,\ytEx) -- (\xr+0.4,0);
    \draw[-latex,decorate,decoration={snake,amplitude=.2mm,segment length=1mm,post length=1mm}] (\xr+0.8,\yqEx) -- (\xr+0.8,\ytEx);
    \draw[-latex,decorate,decoration={snake,amplitude=.2mm,segment length=1mm,post length=1mm}] (\xr+1.0,\ytEx) -- (\xr+1.0,\yfEx);
    \draw[-latex,decorate,decoration={snake,amplitude=.2mm,segment length=1mm,post length=1mm}] (\xr+1.6,\ysEx) -- (\xr+1.6,\yfEx);
    \draw[-latex,decorate,decoration={snake,amplitude=.2mm,segment length=1mm,post length=1mm}] (\xr+1.8,\yfEx) -- (\xr+1.8,0);
    
    \node[anchor=west,font=\footnotesize] at (\xr+0.2,\ytEx+0.5) (g1) {$\gamma_1$};
    \node[anchor=west,font=\footnotesize] at (\xr+0.4,\yfEx+1.4) (g2) {$\gamma_2$};
    \node[anchor=west,font=\footnotesize] at (\xr+0.8,\ytEx+0.5) (g3) {$\gamma_3$};
    \node[anchor=west,font=\footnotesize] at (\xr+1.0,\yfEx+1.4) (g4) {$\gamma_4$};
    \node[anchor=west,font=\footnotesize] at (\xr+1.6,\yfEx+1.4) (g5) {$\gamma_5$};
    \node[anchor=west,font=\footnotesize] at (\xr+1.8,0.6) (g6) {$\gamma_6$};
    
    \node[anchor=west,font=\footnotesize] at (-1,3.2) (eg1){$E_{\gamma_1} = 3.87$~MeV};
    \node[anchor=west,font=\footnotesize] at (-1,2.8) (eg1){$E_{\gamma_2} = 4.24$~MeV};
    \node[anchor=west,font=\footnotesize] at (-1,2.4) (eg1){$E_{\gamma_3} = 1.00$~MeV};
    \node[anchor=west,font=\footnotesize] at (-1,2.0) (eg1){$E_{\gamma_4} = 2.87$~MeV};
    \node[anchor=west,font=\footnotesize] at (-1,1.6) (eg1){$E_{\gamma_5} = 2.75$~MeV};
    \node[anchor=west,font=\footnotesize] at (-1,1.2) (eg1){$E_{\gamma_6} = 1.37$~MeV};
    
	\end{tikzpicture}
    \caption{Decay scheme of $^{24}$Na}\label{schemeNa}
\end{figure}

Scientifically, the perspectives offered by \Na\ are more promising.
The signal in this case would be given by the detection of the $\beta$ with the largest branching ratio
($\beta_2$ in Figure~\ref{schemeNa}) followed by $\gamma_5$ with $2.75$~MeV,
with the corresponding disappearance of $\gamma_6$ with $1.37$~MeV.
For \Na, there are two intrinsic backgrounds, both of which can be easily suppressed.
With reference Figure~\ref{schemeNa}, the parent nucleus can decay via the emission of $\beta_3$.
If $\beta_3$ has enough energy and $\gamma_6$ is absorbed in vicinity of the source,
this could be misinterpreted as the sum of $\beta_2$ and $\gamma_5$.
This type of event can easily be removed with the requirement that $\gamma_5$ has to be
detected sufficiently far from the source.
A second background arises in the main branch if $\beta_2$ has very small energy
and $\gamma_6$ is absorbed next to the source.
This possibility can be rejected setting an upper limit smaller than 
$E_{\gamma_6}$ for the energy deposition next to the source.
The main drawback of \Na\ is its short half-life of about 15~hr,
which necessitates the placement of the detector in the vicinity of the source production site,
as well as the use of a detector technology which allows the repeated source insertion and removal.
On the long term, one could envisage a two stages approach in which \Co\ is first used
to test and improve the technology, followed by a \Na\ phase with more ambitious physics goals.

Finally, a small fraction of the decays proceed through cascades with second photons with much higher energy than the benchmark photons we discuss for both \Co\ and \Na\ ($\gamma_3[\text{Trigger}] +\gamma_4[\text{Miss}]$). This increases the reach to higher $\phi$ masses albeit with lesser sensitivity. This shows up as a kink in Figure~\ref{expreach} .


\subsection{Design Characteristics \label{sec:designcharacteristics}}

\begin{figure}[htbp]
    \centering
    \includegraphics[width=\columnwidth]{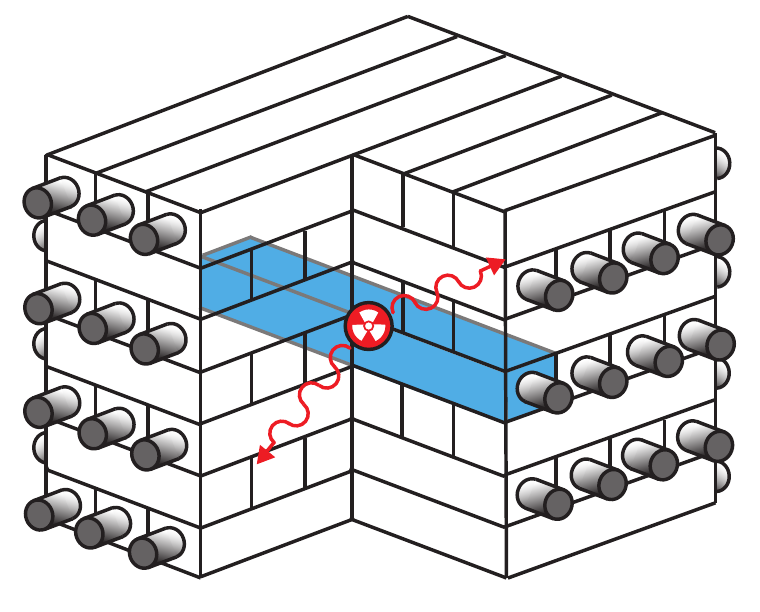}
    \caption{Schematic rendering of the experimental design:
    the scintillator modules are stacked in layers with alternating orientations
    and are coupled to light detectors (grey cylinders) on both ends. The central module, 
        used for triggering on $\beta$ events, is shown in blue and can be made of a 
            different material or size. }
    \label{fig:mockup}
\end{figure}

The required high containment efficiency and time resolution can only be achieved
with liquid or solid scintillators. If a \Na\ source is to be used,
a promising design is that of a stack of solid scintillator modules
with the source as a thin foil at the center, as depicted in
Figure~\ref{fig:mockup}.
To minimize the dead volume, one can substitute the standard reflective foils
with ultra-thin nano-fabricated coatings, for which the technology is readily available.
The detection efficiency can be maximized by coupling 
light detectors (e.g. PMTs or SiPMs) at the two ends of each module.
While some scintillating crystals offer higher light yields ($LY$) of up to 
$6\cdot10^4$~photons/MeV~\cite{StGobain},
their maximum size is limited by crystal growth
technology, and their cost tends to scale up quickly with size.
Plastic scintillators on the other hand can typically sustain a higher count rate
thanks to lower decay times, are cheaper, and can be molded
in almost arbitrary shape and size.
Their main drawback is the light output, limited to $\sim10^4$~photons/MeV~\cite{StGobain}.

We developed a full Geant4~\cite{geant4} 
Monte Carlo (MC) simulation to evaluate
the containment efficiency as a function of the active detector size,
and quantify the importance of the intrinsic \Co\ background described above.
For simplicity, we simulated a cubic active volume of side $l$,
composed of rectangular cuboids of volume $l \times d \times d$
arranged in alternating orientations.

In the following calculations, we assume a LY of $10^4$~photons/MeV (typical
for a plastic scintillator such as BC-404~\cite{StGobain}) and conservatively
scale it down by a factor 3 to account for the self-adsorption and possible
inefficiencies in the light propagation to the detetor,
a $25\%$ Quantum Efficiency ($QE$) for the light detectors,
and an energy resolution given by:
\begin{equation}
    \sigma_E = \frac{ \sqrt{ 3 \cdot LY \cdot QE \cdot E } }{ LY \cdot QE }\quad.
\end{equation}

\section{Event Selection\label{sec:selection}}

As mentioned above, the signal signature is an energy deposition
compatible with that of the $\beta_1$ in the module(s) next to the source, and an energy compatible with that of $\gamma_1$ for \Co\ deposited elsewhere, or with $\beta_2$ and $\gamma_5$ for \Na. In order to mitigate backgrounds, we segment the detector into three regions (see figure \ref{fig:mockup}). First, we have a central module around the source whose purpose is to measure the beta from the source. This module will have a size $\sim$ cm, so that it can completely stop the $\sim$ MeV betas produced by the source. Surrounding this central module, we will have an inner module of thickness $\sim$ 10 cm, corresponding to one radiation length of the expected gammas. The inner modules are surrounded by outer modules that extend to sufficiently many radiation lengths to achieve the necessary containment. 

Our event selection protocol works as follows: we demand that there is an energy deposition in the central module consistent with the initial beta. We then demand that the subsequent gammas deposit all of their energy in the inner modules of the detector. If there is any energy deposited in the outer modules within the $\sim$ ns timing resolution of the experiment or if the gamma ray energy deposited in the inner module is inconsistent with the expected energy, we veto the event. 

This strategy sacrifices $\mathcal{O}\left(1\right)$ of the signal, where the gammas travel a few radiation lengths before scattering or have soft collisions in these inner modules. On the other hand,  this eliminates the need to carefully reconstruct activity that occurs in the outer modules which house most of the volume of the detector. 

\section{Sensitivity and Backgrounds\label{sec:sensitivity}}

To estimate the efficacy of the above signature, we define a region of interest (ROI) $[E_\gamma-n\sigma_E,E_\gamma+n\sigma_E]$ around the energy
of the considered gamma. All processes which can mimic this signature represent possible backgrounds which can hinder a discovery.

To quantify the effect of each background, we compute the $3 \sigma$ discovery
sensitivity as a function of the live time of the measurement $t$
and of the other experimental parameters.
In general, the number of signal events $s$ can be written as:
\begin{equation}
    s = \varepsilon_t \cdot \varepsilon_{ROI} \frac{\Gamma(\phi)}{\Gamma_\gamma}
    \cdot \varepsilon_{MC} \cdot A \cdot t \quad,
\end{equation}
where $\varepsilon_t$ is the trigger efficiency, $\varepsilon_{ROI}$
is the fraction of signal events with energy deposition in the inner modules
falling in the selected ROI, $\varepsilon_{MC}$ is the containment efficiency,
and $A$ is the source activity.
The number of background events is given by:
\begin{equation}
    b = \varepsilon_t \cdot \varepsilon_b \cdot A \cdot t \quad,
\end{equation}
where $\varepsilon_b$ is the probability of a specific background to induce
an event in the ROI.
In all calculations we can safely assume $\varepsilon_t = \varepsilon_{MC} = 1$,
and compute the sensitivity as a function of the exposure $A\cdot t$.
We define the discovery sensitivity as that value of $\Gamma(\phi)/\Gamma_\gamma$
for which an experiment has a $50\%$ probability to measure a positive signal
above background with a significance of at least $3\sigma$.
We compute this following the heuristic counting approach
described in Ref.~\cite{Agostini:2017jim}.

\subsection{Photon Miss}

The first background arises if the gamma under investigation
is not absorbed in the active detector volume.
In our design, this can happen only in the source itself,
or if the gamma escapes undetected.
Therefore, the source must consist of a thin enough foil
to make the self-adsorption negligible, and the detector size must
be such that $\varepsilon_{MC}\geq (1-10^{-A\cdot t})$.
In order to reach a $\hat{g}_p \sim 10^{-7}$, the total detector
size must cover $32$ interaction lengths,
corresponding to $\sim10$~m for BC-404.
Such a high containment represents a major technological
challenge for the proposed design,
as the presence of empty and dead volumes has to be avoided at any cost.
Empty volumes can presumably be avoided using scintillator modules
with non-trivial shapes to avoid direct lines of sight
between the source and the outside world, together with the aforementioned 
thin film reflective coatings.

\subsection{1.33~MeV gamma mimicking 1.17~MeV gamma}

\begin{figure}
    \centering
    \includegraphics[width=\columnwidth]{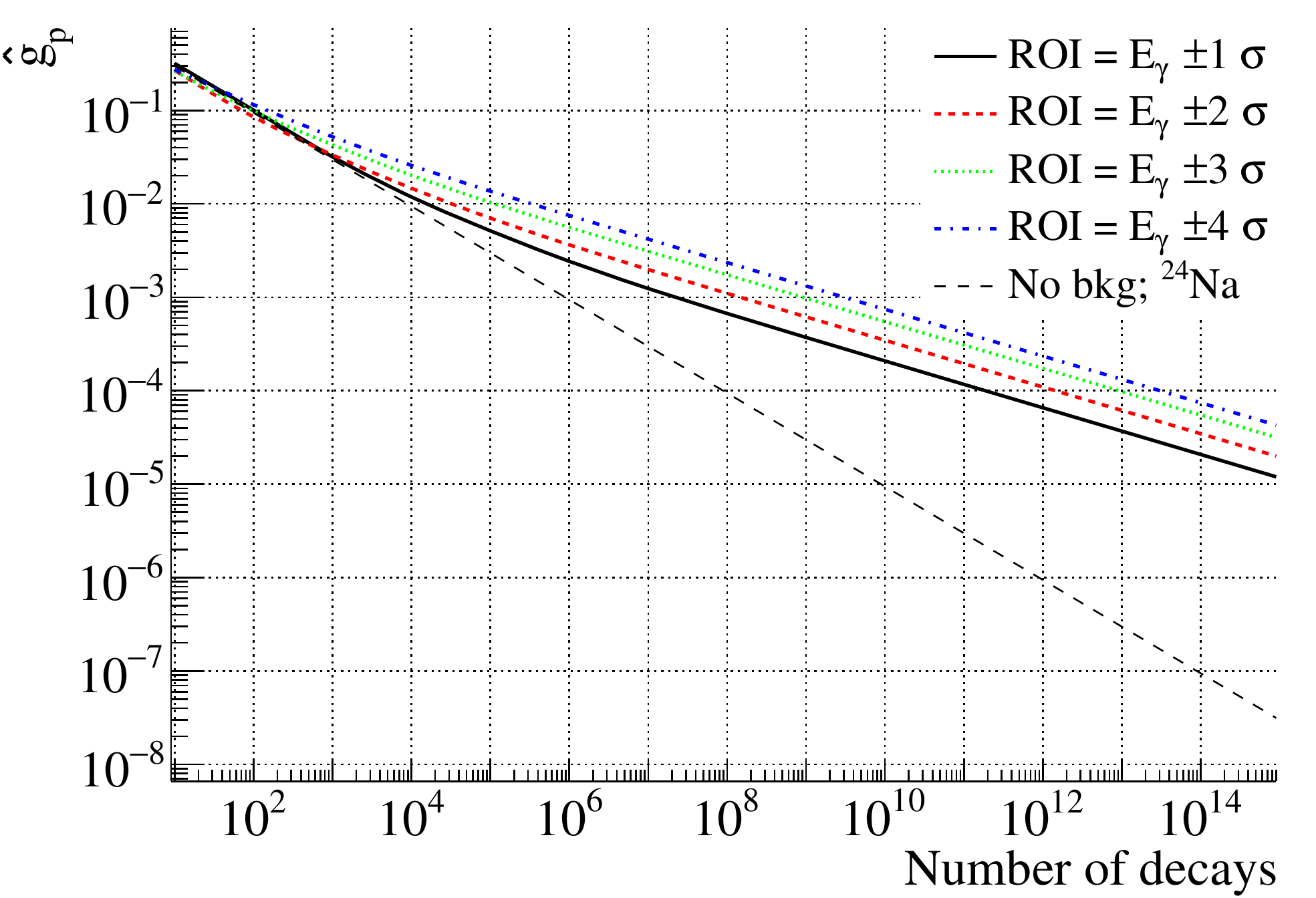}
    \caption{Discovery sensitivity curve at $3 \sigma$ significance
    with \Co\ for different choices of the ROI. The dashed line
    shows the case with no background, or with \Na.}
    \label{fig:sensitivityRes}
\end{figure}

In the \Co\ case, another background is induced by the misreading
of a 1.33~MeV gamma as a 1.17~MeV one. This is possible if the energy resolution
is such that a non-negligible fraction of 1.33~MeV events can fall in the ROI.
It can only occur if the decay follows the weaker \Co\ branch,
and if $E_{\beta_2}<0.32$~MeV. The background efficiency is:
\begin{equation}
\varepsilon_b = 0.0012 \cdot P(E_{\beta_2}<0.32~\text{MeV}) \cdot P(E_{\gamma_2} \in \text{ROI}).
\end{equation}
MC simulations give $P(E_{\beta_2}<0.32~\text{MeV}) \simeq 0.8$,
with a weak dependence on the scintillator module size, $d$.
Using this value, we obtain the discovery sensitivity
as a function of the exposure shown in Figure~\ref{fig:sensitivityRes}.

\subsection{Soft-Compton events}

\begin{figure}
    \centering
    \includegraphics[width=\columnwidth]{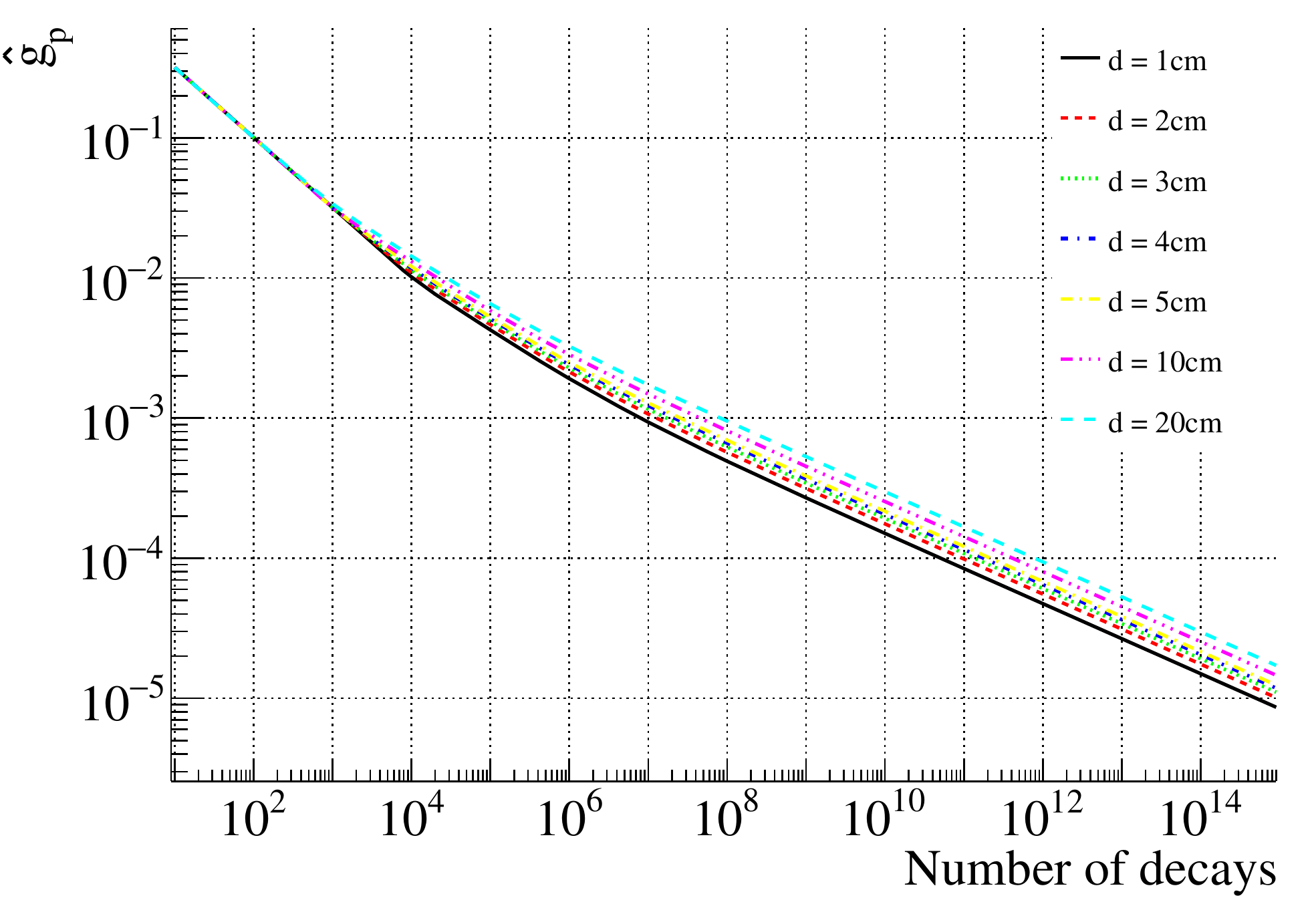}
    \caption{Discovery sensitivity curve at $3 \sigma$ significance
    with \Co\ with the inclusion of the soft-Compton
    for different sizes of the scintillator module.}
    \label{fig:sensitivitySoft}
\end{figure}

For \Co, the weak decay branch causes a second type of background,
if $E_{\beta_2}<0.32$~MeV, and if $\gamma_2$ makes a soft-Compton
scattering in the same scintillator module(s) where $\beta_2$ is detected.
The background efficiency is:
\begin{multline}
    \varepsilon_b = 0.0012 \cdot P(E(\text{central})<0.32~\text{MeV})\\
    \cdot P(E(\text{elsewhere})\in \text{ROI})\quad .
\end{multline}
The probability for these events to happen increases with the size
of the scintillator modules, as shown in Figure~\ref{fig:sensitivitySoft}.
With $l\lesssim 1$~cm this background is small enough
to allow reaching a sensitivity of $\hat{g}_p\sim 10^{-5}$
with one year of live time.
In practice, though this is not illustrated in the cartoon schematic in 
Figure~\ref{fig:mockup} and tested in our simulation, we envisage the use 
of smaller detector modules in vicinity of the source, and of larger modules 
in the outer region. Due to this intrinsic background, \Co\ allows to cover 
only about half of the parameter space comprised by the NM and the SN1978A
excluded regions (Figure~\ref{expreach}).
Advanced event reconstruction algorithms might help identifying
soft Compton events,
but it is not possible to estimate their efficiency
without a dedicated study which is out of the scope of this work.


\subsection{Radioactive Contaminants}

Radioactive contaminants in the detector can be a source of background. For these contaminants to matter, they would have to deposit energy in the central module that is consistent with the initial beta and lead to energy deposition in the inner modules that is consistent with the single  gamma expected in the decay chain. A list of common contaminants and their expected activity in scintillators can be found in \cite{Loach:2016fsk}. The most dangerous contaminant for \Co\, is $^{40}$K, which decays (with a 10 \% branching fraction) to $^{40}$Ar via electron capture, emitting a 1.46 MeV gamma. This contaminant is estimated to occur with an activity $\sim$ mBq  in the $\sim$ kg volume of the inner modules \cite{Loach:2016fsk}. If this gamma undergoes a soft scattering in the central module and subsequently re-scatters in the inner module, it could mimic our signal in \Co. With branching fraction and phase space suppression, this yields worrisome events at a rate of $\sim 10^{-5}$ s$^{-1}$. If the \Co \, experiment is operated at an event rate $\sim$ 10 MHz, branching fractions $\sim 10^{-12}$ can be probed before this background becomes a problem. This background is then comparable to other intrinsic backgrounds in \Co. It is possible to further mitigate this background through a minor sacrifice in the signal. The signal from \Co \, involves a 0.32 MeV beta followed by a 1.17 MeV gamma. By requiring that the beta from the \Co \, decay have less energy than  0.2 MeV, the total energy in the \Co \, decay is resolvably different from the energy produced in $^{40}$K, suppressing this background. There do not appear to be comparably worrisome contaminants for \Na.

\subsection{Cosmics and Neutrons}

The requirement that the event only contain energy deposition in the inner and central modules is a powerful way to discriminate against backgrounds induced by cosmic rays. With the exception of neutrons, cosmic ray activity will lead to a trail of energy deposition in the detector and can be vetoed. Environmental neutrons can potentially cause backgrounds if they get captured in a nucleus, leading to de-excitation through gammas that could be in the right energy range. However, the inner and central modules of the detector are at least $\sim 30$ neutron radiation lengths ($\sim 10$ cm) away from the environment. We thus expect the inner regions to be self-shielded from environmental neutrons. Neutrinos are the only environmental source that can penetrate into the inner modules of the detector. They could be a cause for concern if they are in-elastically absorbed in nuclei, leading to de-excitation through gamma rays. The expected rate for such events in the inner modules of the detector is $\sim 10^{-2}$/yr and is thus not a concern. 

\subsection{Dead Regions}

Solid state scintillators can have dead layers, where energy deposition leads to highly suppressed light emission. These dead layers can be caused due to oxidation, humidity and mechanical damage (see \cite{Nefedov2016} and references therein). In plastic scintillators, the typical thickness of these dead layers is $\sim 1 - 10 \, \mu\text{m}$ \cite{Nefedov2016}. In the outer modules of the detector, whose sole purpose is to identify the existence of a second photon in the event (rather than a precise reconstruction of the energy itself), a dead region would be problematic only if all of the photon's energy was absorbed in that region. However, upon absorption, a MeV gamma would produce an MeV electron which would travel a distance $\sim$ cm before being stopped. Since this is much bigger than the expected thickness of the dead region, we do not expect a background from their existence in the outer modules. 

In the inner module, where energy reconstruction is important, a dead region can be problematic if it causes mis-identification of energy. For example, if \Co \, decays through $\beta_2$ or \Na \, decays through $\beta_3$ and if the energy of the outgoing photon is mis-identified due to the photon going through a dead region, it is possible for these regions to cause backgrounds. In \Na, where the two gammas are well separated in energy, this should not be a problem. In \Co, the gamma would have to lose $\sim$ 100 keV energy in a dead region in order to get confused with the signal photon $\gamma_1$. However, the $\sim$ 100 keV electron produced in this absorption would travel a distance $\sim 100 $ microns before stopping. Since the typical lengths of the dead layers are $\sim 1 - 10$ microns, it does not seem likely that these dead regions would cause a background even for \Co. In the event of a positive signal, since the problem of the dead layers is confined to potentially big ($\sim$ 100 micron) regions in the inner modules, they could be examined to observe the size of such regions \cite{6767156}.

The radiation dose experienced by the central module at an event rate of 10 MHz operating for a year is $\sim 10^4$ Grays, smaller than the dosage ($\gtrsim 10^{5}$ Grays) necessary to cause damage to the module.


\section{Technological challenges and possible improvements\label{sec:challenges}}

The experimental apparatus we propose features several technical challenges, which we believe can potentially be overcome with a careful design of all its components. First, a non-trivial trigger logic is necessary: it must be capable of selecting in real time only signal-like events while handling the high ($\sim$ 10 MHz) data rate. This trigger would also have to allow operation in ``background'' mode to store all detected events for a detailed characterization of the detector response. The design of this trigger is beyond the scope of this work. However, we point out that triggers that need to operate at this high frequency with comparable efficiency are necessary for the LDMX experiment currently being developed at SLAC~\cite{Akesson:2018vlm}. Second, the required containment efficiency has to be achieved while avoiding the presence of significant dead volumes (see estimates above) and minimizing source thickness to prevent self-absorption. Finally, in an experiment using \Na, source production must be possible in the vicinity of the experimental apparatus. Given its high chemical reactivity, a $^{23}$Na compound must be used,
yielding the possibility of parasitic activation of  other molecular components during  neutron irradiation.
A reasonable compound is one which either does not suffer from
parasitic activation, or that does not induce additional background and does not increase the total count rate.

Alternative designs to the one we proposed are also possible.
For example, similar containment efficiencies are achievable with
TPCs or large liquid scintillator experiments. The main disadvantages are limited spatial resolution and difficulties in repeated insertion and removal of the source. One concept worthy of further exploration might be to insert the inner module in a large liquid scintillator, which would serve as the outer module. This design would combine the rapid response necessary in the inner module to handle the high event rate while using the liquid scintillator to identify energy deposition that would trigger an event veto.  Such a detector could potentially be installed in existing large liquid scintillator experiments. In addition, outer modules made out of a liquid scintillator would also mitigate concerns about dead regions in those modules. 

In this work, we made use of a simplified geometry,
of conservative assumptions for the parameters characterizing
the scintillator and the light detectors, and of a trivial
event selection in the analysis of the simulated data.
A large space for improvement is available on both the hardware
and the event reconstruction aspects. Light detectors with up to $35\%$~QE
are available on the market, allowing for a strong improvement
in energy resolution. The geometry and size of the scintillator modules
can be optimized to maximize containment and minimize the intrinsic
\Co\ backgrounds. Finally, event reconstruction algorithms using the information
collected by the single modules can further improve the signal to background ratio.

From a strategic point of view, we propose a multi-stage approach. In a first phase, a \Co\ source can be used to develop, test and prove the performance of the inner modules of the detector where energy reconstruction has to occur with high efficiency at a high rate. If successfully implemented, this phase can lead to a full scale \Co \, experiment that proves containment and background rejection of the outer modules. At a later stage, this setup can be operated with \Na\, at a facility with localized source production. Finally, a positive signal 
could be scrutinized with the use of additional sources inducing more complicated 
signatures (e.g. with multiple gammas) which can be hardly mimicked by external 
backgrounds.

\section{Conclusions\label{sec6}}

Experiments in the intensity/statistics frontier have the ability to probe physics complementary to the high energy parameter space accessed by colliders. Given the existence of weakly-coupled sectors such as those associated with the dark matter and dark energy, there are strong theoretical motivations to search for light, weakly-coupled particles. While a successful implementation of this experiment covers the supernova ``trapping" window, it is clear that there are significant challenges to probe parameters below the range constrained by supernova cooling bounds. Accessing this parameter space will make the experiment sensitive to a wider class of well motivated particles with modifications (other nuclei) sensitive to even the QCD axion and axion-like-particles in a range of parameters that are otherwise hard to constrain. 

To probe this region, the experiment must be sensitive to branching fractions smaller than $10^{-19}$. This involves many obstacles that require additional technology development. Specifically, the required containment efficiency requires a larger detector size and techniques to handle source activity at the GBq level. Event pile up can potentially be mitigated with an optimized choice of geometry wherein the detector modules are at some distance from the source. This would spread out the decays over multiple detection modules, preventing pile up. However, a complex trigger mechanism would be required to do simultaneous event reconstruction for multiple decays. 

On the other hand, it is possible that a more optimized choice of nuclear states could alleviate the statistics necessary to probe these particles. For example, $M_1$ transitions in $^{65}$Ni could be sensitive to the pseudo-scalar axion. Furthermore,  gamma cascades that involve forbidden photon transitions (for e.g. $E_0$ or $M_0$) could offer significant advantages in the search for scalars, axions and axion-like particles. Since these bosons do not carry spin, the branching fraction for axion/axion-like particle production in transitions between forbidden levels would be enhanced. If suitable levels are found, this would decrease the statistics needed to search for these particles as well as decrease the containment efficiency required. 

\section*{Acknowledgements}
The authors believe that the great Mahatma Gandhi would have been thrilled at the prospect of using nuclear physics for peaceful purposes.
The authors sincerely thank M.~Battaglieri, W.~C.~Haxton, Y.~G.~Kolomensky and B.~Schwingenheuer, 
for providing very useful feedback and suggestions. SR was supported in part by the NSF under grants PHY-1638509 and PHY-1507160, the Alfred P. Sloan Foundation grant FG-2016-6193 and the Simons Foundation Award 378243. This work was supported in part by Heising-Simons Foundation grant 2015-038.  HR is
supported in part by the DOE under contract DEAC0205CH11231 and the BFS Fund 27660.
Some of this research was completed in KITP, supported in part by the National Science Foundation under Grant No. NSF PHY-1748958. Part of this work was also completed at Aspen Center for Physics, which is supported by National Science Foundation grant PHY-1607611.

\bibliography{sample}

\end{document}